\documentclass[reprint,amssymb, amsmath, aps, superscriptaddress, showpacs, footinbib, prb]{revtex4-1}
\usepackage{graphicx}
\usepackage{color}
\usepackage[colorlinks,breaklinks,bookmarks=true,citecolor=blue,linkcolor=red,urlcolor=blue]{hyperref}
\newcommand{\mg}{\text{mag}}
\newcommand{\AFM}{\text{AFM}}

\usepackage{color}

\begin{document}

\title{Nearly compensated exchange in the dimer compound callaghanite Cu$_2$Mg$_2$(CO$_3$)(OH)$_6$$\cdot$2H$_2$O}

\author{Stefan Lebernegg}
\email{stefan.l@sbg.at}
\affiliation{Max Planck Institute for Chemical Physics of Solids, N\"{o}thnitzer
Str. 40, 01187 Dresden, Germany}

\author{Alexander A. Tsirlin}
\affiliation{Max Planck Institute for Chemical Physics of Solids, N\"{o}thnitzer
Str. 40, 01187 Dresden, Germany}
\affiliation{National Institute of Chemical Physics and Biophysics, 12618 Tallinn, Estonia}

\author{Oleg Janson}
\affiliation{Max Planck Institute for Chemical Physics of Solids, N\"{o}thnitzer
Str. 40, 01187 Dresden, Germany}
\affiliation{National Institute of Chemical Physics and Biophysics, 12618 Tallinn, Estonia}

\author{Helge Rosner}
\email{Helge.Rosner@cpfs.mpg.de}
\affiliation{Max Planck Institute for Chemical Physics of Solids, N\"{o}thnitzer
Str. 40, 01187 Dresden, Germany}
\date{\today}

\begin{abstract}
A combined theoretical and experimental study of the natural Cu$^{2+}$-mineral callaghanite is \hbox{presented}. Its crystal structure features well separated Cu$_2$(OH)$_6$ structural dimers with weakly bonded carbonate groups and water molecules in between. Susceptibility, field-dependent \hbox{magnetization} and specific heat measurements reveal a compound with a small spin gap of about 7\,K. The observed magnetic properties are well described by a model of isolated antiferromagnetic spin dimers. Possible ferromagnetic interactions between the dimers amount to $-1.5$\,K, at most. Different flavors of electronic structure calculations have been employed to locate the magnetic dimers in the crystal structure, i.e., to determine whether they coincide with the structural dimers or not. Calculations of the coupling between the structural dimers clearly show that magnetic and structural dimers are the same. For the intradimer coupling, however, the computational results confirmed a coupling strength close to zero but the sign of the coupling could not be determined unambiguously. Based on this finding, we then discuss how the reliability of the numerical methods depends on the characteristics of exchange pathways and on structural features of the compound in general. Eventually, we try to provide a minimum coupling strength that is needed for a reliable computational description.
\end{abstract}

\maketitle

\section{Introduction}
Fascinating magnetic behavior, exotic ground states and new quantum phenomena have been discovered over the last years by intensive investigations of spin-1/2 magnetic insulators. Spin-Peierls transitions,\cite{FHC_CuGeO3_spin-Peierls} skyrmions~\cite{skyrmions} and Bose-Einstein condensation of magnons~\cite{natureBose} are only some examples for such phenomena that attracted so much attention since they enable expanding our understanding of the quantum nature of matter and may be relevant for high-tech applications.\cite{ruegg2007,levi2007,skyrmions2} The diversity of magnetic properties observed, in particular in Cu$^{2+}$ compounds is directly related to the huge variety of crystal structures that can be found in this class of materials: The magnetic Cu$^{2+}$-ion and its ligands X typically form (distorted) square-planar CuX$_4$ plaquettes, which can either remain well isolated in the structure or may be linked in various ways to form dimers, chains, planes or even more intricate frameworks. The arrangement and connectivity of the plaquettes as well as their geometric distortion directly determine the strength and type of magnetic exchange interactions and, thus, the macroscopic magnetic properties. Accordingly, the evaluation of individual exchange pathways and the subsequent development of a microscopic magnetic model is essential for understanding the magnetic behavior. This task, however, is far from being trivial. 

As a first step toward the microscopic understanding of magnetic materials, a set of valuable empirical rules has been derived by Goodenough, Kanamori and Anderson (GKA). For cuprates, these rules basically describe how the exchange interactions depend on the Cu--X--Cu bridging angle.\cite{gka1,gka2,gka3} However, a straight forward application of these rules often results in inappropriate microscopic magnetic models, since in general many details, such as covalency, distortions or neighboring anionic groups, play an important role.\cite{braden,gka_lig,cux2} In the last years, the combination of experiments and advanced theoretical methods, such as density functional theory (DFT) or wave-function (WF)-based approaches, has turned out as a reliable strategy for developing microscopic magnetic models in cases where experiment or theory alone may leave ambiguities.\cite{FHC_LiCu2O2_chiT_CpT_NS_wrong_model_paper,*FHC_LiCu2O2_chiT_CpT_NS_wrong_model_comment,*FHC_LiCu2O2_chiT_CpT_NS_wrong_model_reply,cucl2} The ambiguity of describing experimental data with different models usually stems from the large number of independent model parameters which, in turn, allow for similarly good fits of the measured data for several models or parameter regimes. 

In the case of computational methods, ambiguities arise from different approximations for the description of strong electron correlations, for which no feasible general scheme is available so far. The problems are particularly severe in case of weak couplings close to the transition from an antiferromagnetic (AFM) to a ferromagnetic (FM) regime, where subtle features of the crystal structure play a crucial role. However, at least a qualitative evaluation of such couplings is essential because it helps to establish the dimensionality of the system, the relevance of magnetic frustration, and other important details of the magnetic model. Ambiguities in the computational results have been encountered, e.g., for the (Cu$X$)La$M_2$O$_7$ ($X$\,=\,Cl, Br; $M$\,=\,Nb, Ta) family,\cite{cuclnbta} azurite,\cite{azurite} CdVO$_3$\cite{cdvo3} and $\beta$-Cu$_2$V$_2$O$_7$.\cite{betacu2v2o7} In all these cases, the mean-field treatment of Hubbard correlations on top of standard DFT -- the so-called DFT+$U$ method -- was employed. Problems, in particular with respect to the coupling strength, also occurred for other computation techniques that rest upon the admixture of Hartree-Fock exchange to DFT functionals (hybrid functionals)\cite{hybrid1,hybrid2} or an advanced WF-based treatment of electronic correlations.\cite{graaf_li2cuo2,li2cuo2_ins,ddci_2} 

In the present work, we investigate the natural Cu$^{2+}$-mineral callaghanite. It features well isolated Cu$_2$(OH)$_6$ dimers (Fig.~\ref{struct}) with a \hbox{Cu--O--Cu} bridging angle of 96.14$^{\circ}$ (Sec.~\ref{sec:xxstr}) that falls into the range of 95--98$^{\circ}$ where the transition from FM to AFM coupling typically occurs.\cite{braden} This behavior follows the empirical GKA rules, which require that FM couplings prevail for bridging angles close to 90$^{\circ}$, where the usually dominating AFM second-order contributions vanish for symmetry reasons. Accordingly, callaghanite is a good candidate for nearly compensated FM and AFM exchange contributions at the bridging angle of about $96^{\circ}$. This expectation is actually supported by our magnetization and specific heat measurements (see Sec.~\ref{sec:experiment}) revealing a quantum paramagnetic behavior and very weak exchange couplings with an absolute strength of below 10\,K. From an experimental point of view, this small energy scale renders the compound also a good candidate for observing interesting effects, such as quantum phase transitions under pressure. On the theoretical side, on which we will focus in the present study, this mineral represents an ideal system for testing the accuracy and reliability of different computational methods applied to strongly correlated compounds. DFT+$U$, the PBE0 hybrid functional and WF-based multi-reference methods will be employed for calculating intra- and interdimer couplings, and the results will be compared and evaluated with respect to the experimental data.

The paper is organized as follows: Experimental and theoretical methods are described in Sec.~\ref{sec:methods}. Details of the crystal structure of callaghanite and their relation to the exchange couplings are discussed in Sec.~\ref{sec:xxstr}. The exchange couplings calculated with the different numerical methods will be presented in Sec.~\ref{sec:magexc}. Sec~\ref{sec:experiment} contains all experimental results. A detailed discussion and summary will be given in Secs.~\ref{sec:discussion} and~\ref{sec:summary}, respectively.

\section{Methods}
\label{sec:methods}
The experimental part of this work was done on a natural sample (Fig.~\ref{struct}) of callaghanite from the Premiers Chemical Mine, Gabbs, Nye Co., Nevada, USA. The sample quality was first checked by laboratory  powder x-ray diffraction (XRD) (Huber G670 Guinier camera, CuK$_{\alpha\,1}$ radiation, ImagePlate detector, $2\theta\,=\,3-100^{\circ}$ angle range). Powder XRD patterns down to a temperature of 10\,K were collected at the ID31 beamline of the European Synchrotron Radiation Facility (ESRF, Grenoble) at a wavelength of about 0.4\,\r{A}. 

Magnetization measurements were done on a Quantum Design (QD) SQUID MPMS up to 5\,T and a QD PPMS vibrating sample magnetometer up to 14\,T in a temperature range of 1.2--400\,K. Heat capacity data were acquired by relaxation technique with a QD PPMS in fields up to 8\,T.

The electronic and magnetic structure calculations within DFT were performed with the full-potential local-orbital code \texttt{fplo9.07-41}\cite{fplo} as well as with the Vienna Ab initio Simulation Package (\texttt{VASP5.2}).\cite{vasp1,*vasp2} The first code was used in combination with the local density approximation (LDA),\cite{pw92} generalized gradient approximation (GGA)\cite{pbe96} and the DFT+$U$ method.\cite{lsdau1,lsdau2} Calculations using the PBE0 hybrid functional\cite{pbe0a,*pbe0b} were carried out with \texttt{VASP5.2}. A 4$\times$4$\times$4 $k$-mesh was employed for LDA and GGA runs while supercells used for DFT+$U$ and PBE0 calculations were computed for about 20 $k$-points in the irreducible wedge of the Brillouin zone. The convergence with respect to the $k$-mesh was carefully checked.

The hydrogen positions, which are essential for the evaluation of the exchange couplings~\cite{clinoclase} but not yet determined experimentally,\cite{xxstr} were obtained by a relaxation of the H atomic parameters with respect to the total energy within GGA.\cite{supplement} Such a procedure was recently proven to provide sufficiently accurate H-positions for cuprates.\cite{malachite} 

The exchange coupling constants were calculated in two different ways within DFT. One strategy starts from an LDA band structure. Since LDA does not properly account for strong electron correlations, it typically yields a spurious metallic ground state. The half-filled bands at the Fermi level, however, allow identifying the crucial exchange pathways and are sufficient for the calculation of the low-energy part of the magnetic excitation spectrum. Projecting these bands onto a tight-binding (TB) model, we obtain the transfer integrals $t_{ij}$, which were calculated as off-diagonal Hamiltonian matrix-elements between Cu-centered Wannier functions. Next, the TB-model was projected onto a single-band Hubbard model that accounts for the strong electron correlations by including the effective onsite Coulomb repulsion, $U_{\text{eff}}$, pertaining to the mixed Cu-O Wannier functions, 
$\hat{H}=\hat{H}_{TB}+U_{\text{eff}}\sum_{i}\hat{n}_{i\uparrow}\hat{n}_{i\downarrow}$. Subsequently, the Hubbard model is mapped onto a Heisenberg model 
\begin{equation} 
\hat{H}=\sum_{\left\langle ij\right\rangle}J_{ij}\hat{S_{i}}\cdot\hat{S_{j}}
\end{equation}
This is justified for $t_{ij}\ll U_{\text{eff}}$ and half-filling, as realized in callaghanite (Table~\ref{tJ}). The AFM contributions to the exchange constants $J_{ij}$ are then obtained in second order as $J_{ij}^{\text{AFM}}=4t_{ij}^2/U_{\text{eff}}$, where we used $U_{\text{eff}}=4.5$\,eV according to our previous studies on cuprates.\cite{diaboleite,malachite} 

Alternatively, the full exchange constants, containing also the FM contributions $J_{ij}=J_{ij}^{\text{AFM}}+J_{ij}^{\text{FM}}$, are obtained by including the electron correlations into the numerical procedure. For DFT+$U$ and PBE0, the $J_{ij}$ are calculated as difference of total energies of various collinear (broken-symmetry) spin states which are projected onto a classical Heisenberg model~\cite{bs2}$^,$\footnote{Spin states from DFT calculations are, in general, not eigenstates of the Heisenberg Hamiltonian due to their single Slater determinant character. Thus, they have to be projected onto a classical Heisenberg or an Ising Hamiltonian containing only the $S_z$ components of the spin operators.\cite{bs2}} where we followed the procedure proposed by Xiang et al.\cite{j_xiang} For DFT+$U$ calculations, LSDA+$U$ and GGA+$U$ were used in combination with around mean field (AMF) as well as fully localized limit (FLL) double-counting corrections (DCCs) as implemented in \texttt{fplo9.07-41}. The onsite Coulomb repulsion of the Cu(3$d$) orbitals, $U_d$, was varied between 5.5--8.0\,eV and 7.0--11.0\,eV for AMF and FLL DCCs, respectively, covering the ranges of $U_d=6.5\pm0.5$\,eV and $U_d=8.5\pm1.0$\,eV that are typically used for these two types of DCCs as implemented in \texttt{fplo9.07-41}.\cite{malachite, clinoclase,cux2,diaboleite} The onsite Hunds exchange, $J_d$, was fixed to 1.0\,eV. 

Additionally, we evaluated the intradimer exchange constant $J$ with WF-based methods, which allow for an in principle parameter-free treatment of electron correlations. The calculations were all done in a scalar-relativistic mode using the \texttt{Orca2.9} code.\cite{orca1,orca2} Since the application of WF-based methods is restricted to finite systems with a limited number of atoms, $J$ is calculated for an isolated [Cu$_2$(OH)$_6$]$^{2-}$ cluster. The crystal potential is modeled by embedding the cluster into total ion potentials (TIPs),\cite{tip1} representing the nearest-neighbor Cu$^{2+}$ and Mg$^{2+}$ cations explicitly, and a large array of about 30000 point charges. The point charges were optimized so that PBE0 and unrestricted Hartree-Fock (UHF) cluster results for $J$ agree with those from periodic calculations performed with \texttt{VASP5.2}. In order to reduce the number of electrons in the calculations, the inner 10 electrons (Ne core) of Cu were simulated with a Stuttgart-Dresden effective core potential (ECP).\cite{ecp_cu1,*ecp_cu2} The following basis sets were used for the calculations: def2-TZVPP basis for Cu valence electrons,\cite{tzvpp} aug-cc-PVTZ for oxygen~\cite{ccpvtz} and a simple 3-21G basis for hydrogen.\cite{sto3g} The basis set convergence was carefully checked with PBE0 as well as with $N$-electron valence state perturbation theory (NEVPT2).\cite{nevpt2} Starting from a broken-symmetry LDA-WF, we performed complete active space self-consistent field (CASSCF) calculations with a minimum active space, including the two unpaired electrons in two orbitals. Dynamical correlations were subsequently included by the difference dedicated configuration interaction (DDCI3) method.\cite{ddci_1,*ddci_2} Owing to the fact that DDCI was designed for computing energy differences, it represents one of the most accurate schemes for calculating exchange constants. All numerical calculations are performed for the room-temperature crystal structure. Effects introduced by temperature are discussed in Sec.~\ref{sec:discussion}.

Quantum Monte Carlo (QMC) simulations and exact diagonalization (ED) studies
were performed using the software package \textsc{alps-1.3}.\cite{alps1.3,
*alps2.0} The temperature dependency of the magnetic susceptibility was simulated
using the code \textsc{loop}.\cite{loop} We used finite chains (rings) of up to
$N$\,=\,80 spins $S$\,=\,1/2 with periodic boundary conditions. 50\,000 and
500\,000 sweeps were used, respectively, for and after thermalization.
The magnetic specific heat was simulated using ED on finite chains (rings)
$N$\,=\,16 spins of $S$\,=\,1/2.\cite{[{In the $S$\,=\,1/2 alternating
Heisenberg chain model, finite size effects are generally small except for two
cases, irrelevant for this study: (i) nearly uniform chain limit, when both
exchanges are AFM and (ii) the case of a very large FM coupling, see }] [{}]
hida1992} Magnetization curves were simulated using the QMC code \textsc{dirloop\_sse}\cite{dirloop_sse}. We employed chains of $N$\,=\,80
spins and used 10\,000 sweeps for thermalization and 100\,000 sweeps after thermalization.

The thermodynamic behavior of an isolated Heisenberg dimer can be evaluated
analytically. The reduced magnetic susceptibility $\chi^{*}$ per spin is given by
the expression
\begin{equation}
\label{E-chith}
\chi^*(T,h)=\frac{\eta\left(\eta^2\left(\epsilon+1\right) +4\eta +  \epsilon + 1 \right)}{2T\left(1 + \eta \left(1 + \epsilon + \eta\right)\right)^2},
\end{equation}
where $\eta=\exp{(h/T)}$, $ \epsilon=\exp{(1/T)}$, $T$ is the absolute temperature, and $h$
is the uniform magnetic field. For the zero-field case, the expression readily
reduces to
\begin{equation}
\label{E-chit}
\chi^*(T,h=0)=\frac{1}{T\left(3 + \exp{(1/T)}\right)}.
\end{equation}
The magnetic specific heat $C_p^*$ per spin is given by
\begin{equation}
\label{E-cpth}
C_p^*(T,h)=\frac{\eta^3 \left(\epsilon (h\!-\!1)^2+h^2\right)\!+\!\eta^2 \left(\epsilon\!+\!4 h^2\right)\!+\!\eta \left(\epsilon
(h\!+\!1)^2\!+\!h^2\right)}{2T^2 \left(1+\eta \left(1+\epsilon+\eta\right)\right)^2}.
\end{equation}
Again, in the zero-field case, the expression can be simplified:
\begin{equation}
\label{E-cpt}
C_p^*(T,h=0)=\frac{3\exp{(1/T)}}{2T^2\left(3+\exp{(1/T)}\right)^2}.
\end{equation}
Finally, the uniform magnetization $M^*$ (per spin) as a function of the magnetic field
at finite temperature is given by
\begin{equation}
\label{E-mh}
M^*(T,h)=\frac{\eta^2-1}{2\left(1+\eta\left(1+\epsilon+\eta\right)\right)}.
\end{equation}

\section{Crystal structure}
\label{sec:xxstr}

\begin{figure*}[tbp] \includegraphics[width=17.1cm]{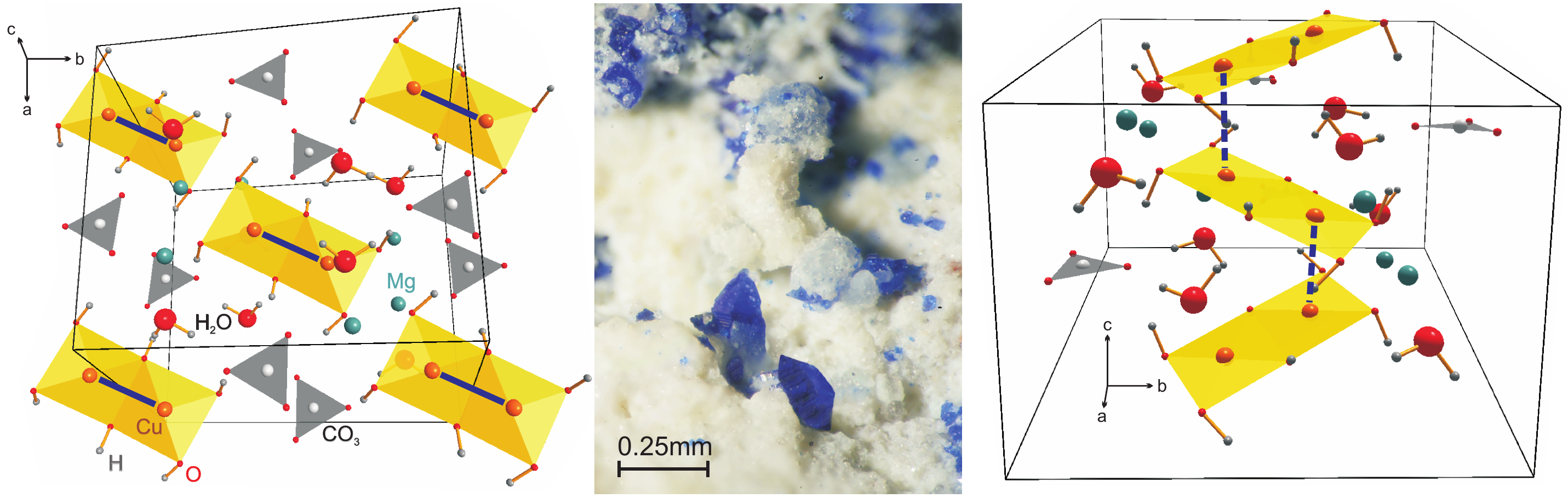}
\caption{\label{struct}(Color online) In the left and right panels the crystal structure of callaghanite is displayed with the Cu$_2$(OH)$_6$ dimers shown in yellow and the CO$_3$ groups in grey color, respectively. Solid and dashed blue bars indicate the intradimer coupling $J$ and the strongest interdimer coupling $J'$, respectively. The central panel shows blue, bipyramidal callaghanite crystals together with colorless hydromagnesite (Mg$_5$(CO$_3$)$_4$(OH)$_2$$\cdot$4H$_2$O) from the Premiers Chemical Mine, Gabbs, Nye Co., Nevada, USA.}
\end{figure*}

Callaghanite crystallizes in the monoclinic space group $C2/c$ and features isolated Cu$_2$(OH)$_6$ dimers (Fig.~\ref{struct}).~\cite{xxstr} Mg$^{2+}$-ions bond to the terminal oxygen of these dimers, while carbonate groups as well as water molecules weakly interact with the dimers via hydrogen bridges. The \hbox{Cu--O--Cu} bridging angle within the dimers amounts to 96.14$^{\circ}$ in the room temperature structure, thus, falling into the range of 95--98$^{\circ}$ where a compensation of FM and AFM contributions to the isotropic exchange coupling typically occurs.\cite{braden,cux2} The distances between Cu and the two bridging oxygens are slightly different, 1.93\,\r{A} and 1.96\,\r{A}, and the dimer features a slight twisting of about 3$^{\circ}$. Both structural details, though being small, may have an effect on the intradimer coupling $J$.\cite{ruiz98,leberneggCCA,buckling} Despite the slight distortions, the dimer retains inversion symmetry that forbids anisotropic Dzyaloshinskii-Moriya interactions.

For the computation of exchange coupling constants and the development of a microscopic magnetic model, accurate crystallographic data are, thus, of crucial importance. The crystallographic data of Ref.~\onlinecite{xxstr} were collected at ambient conditions. Since the magnetic effects in callaghanite occur at lowest temperatures (see Sec.~\ref{sec:experiment}), we thoroughly checked temperature effects on the crystal structure in a range from 300\,K to 10\,K.\cite{supplement} However, we could not find any significant changes of the structural parameters over the whole temperature range. Slight differences between our data and the single-crystal data of Ref.~\onlinecite{xxstr} arise most likely from the use of powder material in our measurements. According to the higher accuracy that can be gained with single crystals, we use the respective data set for our calculations of the electronic structure of callaghanite.  

Of crucial importance for $J$ are, furthermore, the distances and, in particular, the bond angles of the hydrogen atoms bonded to the bridging oxygen.~\cite{clinoclase,ruiz97_2} We obtained the H-positions, undetermined so far, by using a GGA functional and optimizing  their atomic parameters~\cite{supplement} starting from the crystal structure of Ref.~\onlinecite{xxstr}. This procedure was recently proved to be sufficiently accurate for the evaluation of microscopic magnetic models.\cite{malachite} The resulting distance between H and the bridging oxygen is 0.99\,\r{A} and the O--H bond is rotated out of the dimer plane by about 50$^{\circ}$. Note that this large out-of-plane angle should strongly reduce the intradimer transfer $t$ and the corresponding exchange $J^{\text{AFM}}$. Eventually, this drives $J$ toward a FM coupling\cite{clinoclase,leberneggCCA2} (see Sec.~\ref{tb}) and puts callaghanite close to the regime of a complete compensation of the FM and AFM exchanges.

\section{Electronic structure and magnetic exchange couplings}
\label{sec:magexc}

\subsection{\label{tb}LDA results and the tight-binding model}

\begin{figure}[tb] \includegraphics[width=8.6cm]{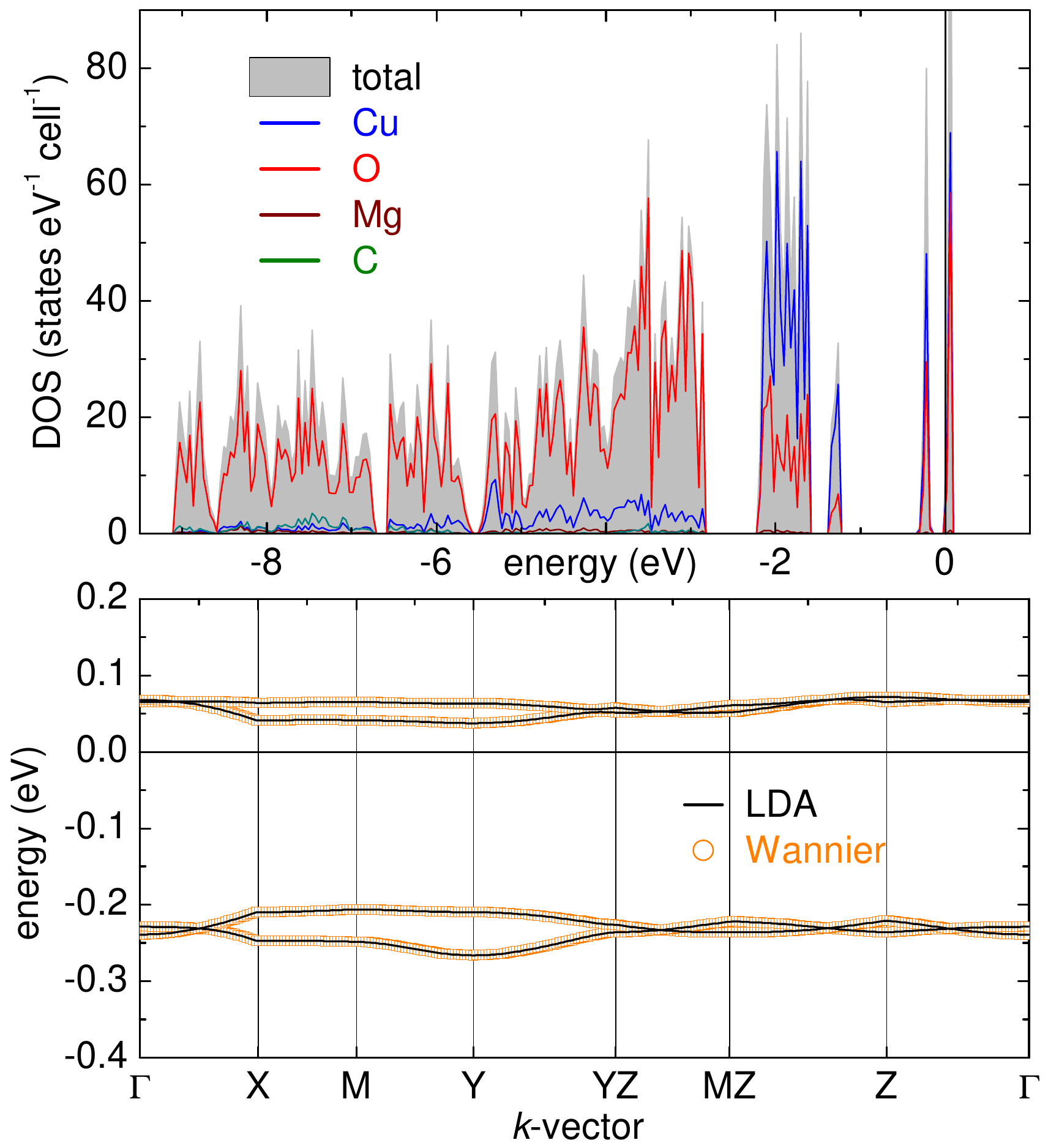}
\caption{\label{lda}(Color online) The top panel shows the total and partial density of states (DOS) from LDA calculations. In the lower panel, the four LDA bands around the Fermi level ($E_F=0$) are shown. "Wannier" denotes bands calculated with Cu-centered Wannier functions. The $k$-points are defined as follows: $\Gamma=(000)$, X$=(\frac{\pi}{a}00)$, Y$=(0\frac{\pi}{b}0)$, Z$=(00\frac{\pi}{c})$, M$=(\frac{\pi}{a}\frac{\pi}{b}0)$, MZ$=(\frac{\pi}{a}\frac{\pi}{b}\frac{\pi}{c})$, YZ$=(0\frac{\pi}{b}\frac{\pi}{c})$.}
\end{figure}

Our LDA calculations yield a broad valence band complex of about 9\,eV width (Fig.~\ref{lda}) as typically observed in cuprates.\cite{clinoclase} The bands between $-9$ and $-3$\,eV with a predominant O(2$p$) character consist of bonding Cu-O $pd\sigma$ and $pd\pi$ states as well as nonbonding O states. The block between $-2.2$ and $-1.5$ arises from antibonding Cu-O $pd\pi$* states. According to the Jahn-Teller distortion and the resulting nearly square-planar coordination of Cu$^{2+}$, the $pd\sigma$* bands are split: With respect to a local coordinate system, where the $x$-axis is chosen as one of the Cu-O bonds and the $z$-axis perpendicular to the plaquette plane, the isolated $pd\sigma$* complex at about $-1.3$\,eV can be described as being predominantly of Cu($3d_{z^2-r^2}$) character while the set of four antibonding bands close to the Fermi level essentially belongs to Cu($3d_{x^2-y^2}$) (see supplemental material~\cite{supplement}). Owing to the very weak dispersion of these four bands, the splitting between the occupied and unoccupied bands $\Delta E$ at the center of the Brillouin zone, $\Gamma$, can be interpreted in terms of the molecular orbital (MO) picture for superexchange presented by Hay et al. (Ref.~\onlinecite{hth75}). They showed that the intradimer transfer integral $t$ is one half of the energy gap between the highest occupied and lowest unoccupied MOs. Accordingly, we can estimate $|t|$ as $|t| \approx \Delta E/2\approx150$\,meV. The weak band dispersion is a result of the isolated character of the Cu$_2$(OH)$_6$ dimers that impedes electrons from being transferred between the neighboring dimers, i.e. all types of interdimer transfers $t_{ij}'$ are small. 

According to a simple TB model, we estimate an effective inderdimer transfer $|t'_{\text{eff}}|$ as $\Delta E_{\text{max}} \approx 4\cdot|t'_{\text{eff}}| \approx 14$\,meV, where $\Delta E_{\text{max}}$ is the maximum splitting of the two occupied bands. Since $|t|\gg|t'_{\text{eff}}|$, even LDA yields an insulating ground state, however, with an energy gap that is an order of magnitude too small to account for the blue color of callaghanite crystals. 

For calculating accurate transfer integrals, we project the four bands onto local Cu(3$d_{x^2-y^2}$) orbitals and obtain four Cu-centered Wannier functions (corresponding to four Cu$^{2+}$ per unit cell) perfectly reproducing the LDA bands. The corresponding transfer integrals $t_{ij}$, which nicely agree with our simple estimate, are given in Table~\ref{tJ}.

\begin{table}[tbp]
\begin{ruledtabular}
\caption{\label{tJ} 
The transfer integrals $t_{ij}$ (obtained from Cu-centered Wannier functions) and the AFM contribution to the exchange constants $J^{\text{AFM}}_{ij}=4t_{ij}^2/U_{\text{eff}}$, where $U_{\text{eff}}$\,=\,4.5\,eV. }
\begin{tabular}{c c c c}
   &  Cu-Cu distance (\r{A}) & $t_{ij}$ (meV) & $J^{\text{AFM}}_{ij}$ (K) \\ \hline
$J$     & 2.89 & 141 & 205    \\ 
$J'$    & 3.22 & $-16$ &   2.6    \\
\end{tabular}
\end{ruledtabular}
\end{table}

The intradimer transfer $t$ is of sizable strength with respect to the small bridging angle of 96.1$^{\circ}$ and leads to a strong AFM contribution $J^{\text{AFM}}$ of 230\,K. However, similar to the situation in clinoclase and in many other Cu$^{2+}$ compounds,\cite{clinoclase} a strong FM contribution is expected as well. Another transfer integral $t'$ operates between the dimers along the $c$-axis (Fig.~\ref{struct}) but is one order of magnitude smaller than the intradimer $t$. There is also a large number of further interdimer transfer integrals, $t_{ij}$, which are, however, all below 5\,meV and are thus expected to play a minor role for the magnetic properties of callaghanite (since the related exchange integrals depend quadratically in the $t_{ij}$'s).

In Sec.~\ref{sec:xxstr}, we emphasized the importance of the angle between hydrogen bonded to the bridging O of the dimer and the dimer plane, and claimed that a large angle leads to a reduction of the transfer integral. In the case of callaghanite, this angle amounts to about 50$^{\circ}$, so that for a planar arrangement of H a considerably larger $t$ can be expected. Indeed, by fixing the O--H bonding distance and rotating H into the dimer plane we obtain an intradimer transfer $t$ of 240\,meV, i.e. increased by about 70\%. This entails an enormous increase in $J^{\text{AFM}}$ by 400\,K and would result in strong AFM coupling within the structural dimers, in contrast to the experimental results (Sec.~\ref{sec:experiment}). This estimate again demonstrates the crucial role that the H-positions play for the exchange couplings and magnetic properties of Cu$^{2+}$ compounds.

\subsection{\label{ldau}DFT+$U$ results}
As explained in section~\ref{sec:methods}, the full couplings containing AFM and FM contributions can be obtained by calculating total energies in a self-consistent procedure that accounts approximately for strong correlation effects. First, we present results from the DFT+$U$ method, which in recent years was extensively used for the evaluation of magnetic parameters in insulators.\cite{j_xiang,cubr2,lsdau2,dftu_blaha,azurite,clinoclase,dftu_kom} Previous experience has shown that the resulting exchange couplings $J_{ij}$ are quite sensitive to details of the computational procedure. While the influence of the correlation parameter $U_d$ can be easily understood from the well-known $J_{ij}^{\AFM}\propto 1/U$ expression, the effects of the DFT functional (LDA vs. GGA) and the DCC are more subtle. 

\begin{figure}[tb] \includegraphics[width=8.6cm]{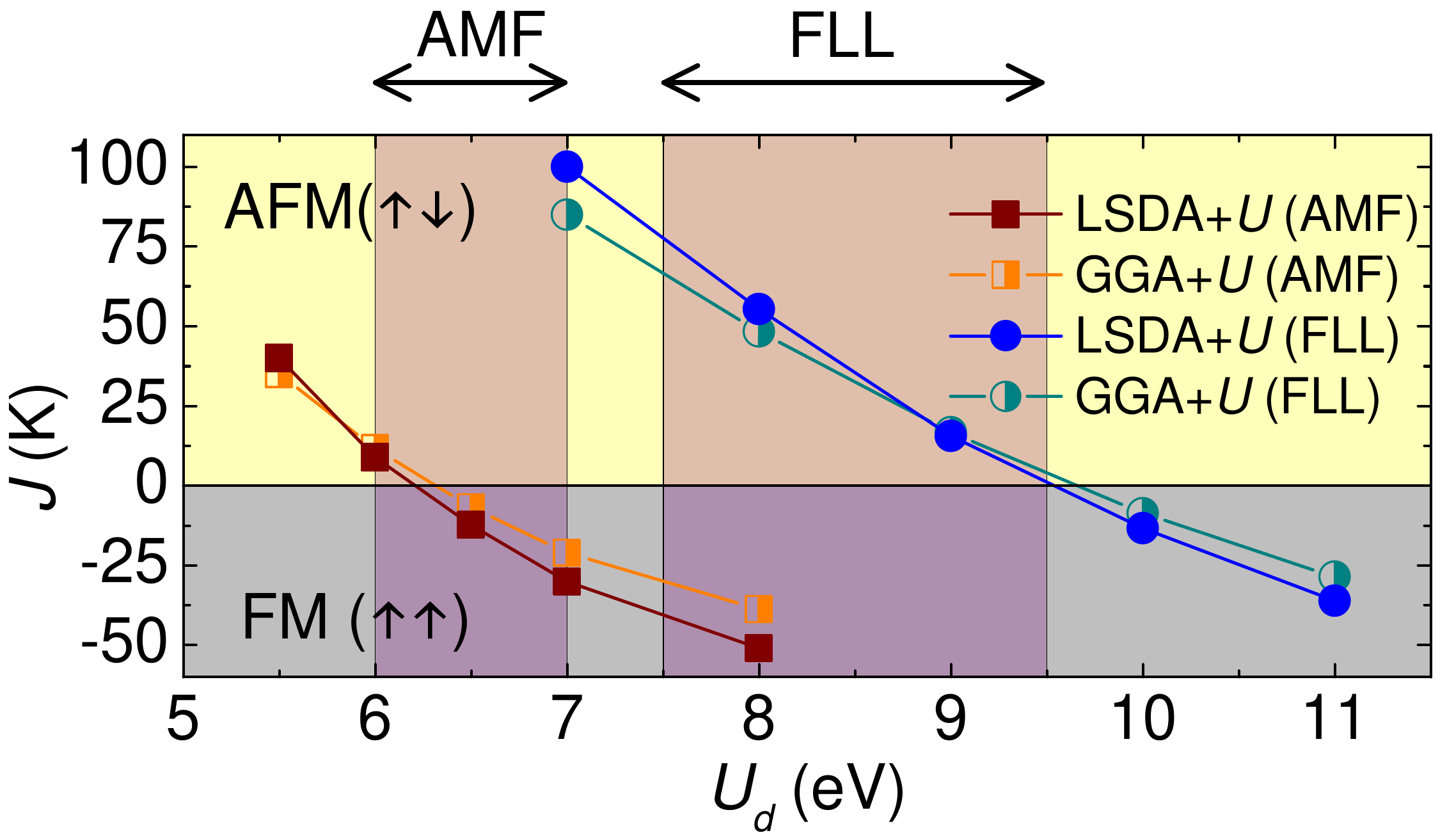}
\caption{\label{lsdau}(Color online) The intradimer exchange constant $J$ calculated with LSDA+$U$ and GGA+$U$ as a function of the parameter $U_d$. AMF and FLL denote around mean field and fully localized limit double counting corrections, respectively. The violet shaded areas indicate the typical ranges of $U_d=6.5\pm0.5$ and $U_d=8.5\pm1.0$ for AMF and FLL DCCs, respectively.\cite{malachite, clinoclase,cux2,diaboleite}}
\end{figure}

In Fig.~\ref{lsdau}, we show the intradimer coupling $J$ calculated for different flavors of DFT+$U$ and with different $U_d$ parameters. As expected, $J$ is drastically reduced by FM contributions. However, depending on the choice of $U_d$, $J$ is either FM or AFM, where FLL seems to favor slightly the AFM side, while AMF somewhat prefers the FM side. Differences between the two types of DCCs have also been observed for other compounds,\cite{azurite,betacu2v2o7} but in the case of callaghanite the situation is dire, because no conclusive information about the sign of $J$ can be obtained.

In contrast to the intradimer coupling $J$, the coupling $J'$ shows only weak dependence on the DCC and $U_d$ where we get $1.6\pm 0.1$\,K and $2.0\pm 0.2$\,K for AMF($U_d=6.5\pm0.5$\,eV) and FLL ($U_d=8.5\pm1.0$\,eV) DCCs, respectively. These results agree well with the TB-analysis also rendering the interdimer coupling quite small. The fact that some couplings are less sensitive to $U_d$ and other details of the computational procedure has been observed for many other compounds as well and will be discussed in detail in Sec.~\ref{sec:discussion}. Regardless of the problems we encountered for the computation of $J$, these results for $J'$ provide clear evidence that magnetic and structural dimers in callaghanite are the same, because experimentally the coupling within the magnetic dimers is $J\approx7$\,K (see Sec.~\ref{sec:experiment}) and definitely exceeds our computational estimates for $J'$.

\subsection{\label{hybrid}Hybrid functionals}
\label{ssec:hybrids}
An alternative method for calculating full $J_{ij}$ in periodic structures is based on the hybrid functionals which are, however, computationally considerably more demanding than DFT+$U$ due to their Hartree-Fock-like exchange-term. We apply here the PBE0 functional~\cite{pbe0} containing a fixed 25\%-admixture of exact exchange. This 25\% fraction was determined on the basis of universal physical constraints, so that PBE0 might be called a nonempirical hybrid functional.\cite{pbe0} Thus, we expect it to be less ambiguous than other hybrids, such as B3LYP that contains purely empirical parameters and is known to severely overestimate the exchange couplings.\cite{b3lyp_acc1,hybrid2} 

According to the unambiguous results for $J'$ that we have obtained with DFT+$U$, we restrict the PBE0 calculations to the intradimer exchange and get a ferromagnetic coupling of $J=-25$\,K. Up to now, only very few studies considered PBE0 for calculating exchange constants in solids, thus, not too much can be said about the accuracy of this method when applied to weak magnetic exchange. In a recent work on CuO,\cite{cuo_p} it is claimed that this functional provides accurate exchange constants, however, even in this study the good agreement is limited to the strong coupling while a weak coupling of about 60\,K is underestimated by almost 50\%(30\,K). 

In order to get more reference data, we used PBE0 to calculate the structurally similar exchange couplings for the dimer compound SrCu$_2$(BO$_3$)$_2$ and for Li$_2$CuO$_2$ featuring chains of edgesharing CuO$_4$ plaquettes. These systems were chosen since the leading exchange constants are precisely known from experiments and because of structural similarities to callaghanite. For the first compound, PBE0 yields $J=128$\,K where the susceptibility measurements provided $J=85$\,K.\cite{srcu2bo32_2003} In Li$_2$CuO$_2$ the nearest neighbor coupling is calculated as $J_1=-283$\,K while inelastic neutron scattering data have been fitted with $J_1=-228$\,K.\cite{li2cuo2_ins} Accordingly, the computed values exceed the experimental data by about 25\% with absolute deviations of more than 55\,K for the presented examples. With respect to these inaccuracies of PBE0, even the sign of $J$, when determined computationally, appears to be questionable in the case of callaghanite.

\subsection{\label{qc}Wave-function based methods}

In contrast to DFT, WF-based methods allow for an in principle parameter-free treatment of electron correlations. They also provide direct access to pure spin states and, thus, do not suffer from possible errors introduced by the broken symmetry formalism~\cite{bs_ddci} used for calculating $J_{ij}$ within DFT. Additionally, published exchange constants from multi-reference CI (MRCI) calculations and its truncations were often in stunning agreement with the experimental data.\cite{hybrid2,ddci_acc2,chem_rev} The major drawback of these methods, however, is their restriction to a small number of atoms entailing the construction of finite cluster models for solids. These models have to be properly embedded to account for the full crystal potential. Since results crucially depend on the cluster choice and embedding, the cluster construction is a nontrivial step for which different strategies have been developed.\cite{emb1,*emb2,*emb3,*emb4} 

In the present study, we focus on the computation of the intradimer coupling $J$, for the same reasons as those given for PBE0, and also because relevant clusters require smaller number of atoms and possess higher symmetry than those for $J'$. Considering the isolated character of the dimers in the crystal structure and their point-group symmetry $C_i$ that makes calculations rather elaborate, we restrict the cluster to a [Cu$_2$(OH)$_6$]$^{2-}$ dimer embedded into TIPs and point charges. The point charges and the charge of the TIPs were optimized so that PBE0 and UHF results for $J$ agree with those from periodic calculations performed in \texttt{VASP5.2}. For the optimized embedding, we obtain $-22$\,K (PBE0) and $-50$\,K (UHF) for the cluster compared with $-25$\,K (PBE0) and $-48$\,K (UHF) for the periodic model. In studies on other compounds, only UHF data were used for comparison.\cite{graaf_li2cuo2,uhf_compare} However, we observed that the UHF results are rather robust with respect to changes in the embedding,\footnote{This problem might be of minor importance if the cluster contains several exchange couplings.} while the PBE0 data are very sensitive, thus, rendering the comparison of the PBE0 exchange constants a more appropriate tool for fine-tuning the embedding.   

Starting from the LDA-WF, we perform a state-averaged CASSCF calculation. The CAS is spanned by the highest occupied (HOMO) and the lowest unoccupied molecular orbital (LUMO) which are both of antibonding character with dominating Cu($3d_{x^2-y^2}$ and O($2p$) contributions. Thus, the CAS comprises the two unpaired electrons of the two Cu$^{2+}$ ions and their singly occupied $3d$-orbitals. Such a minimum CAS is known to be sufficient for calculating exchange coupling constants with MRCI methods.\cite{chem_rev} The molecular orbitals in the CAS are in fact the molecule pendants of the LDA bands around the Fermi level in our periodic calculations (see Fig.~\ref{lda} and supplemental material~\cite{supplement}). The CASSCF calculation based on the minimum CAS yields an FM intradimer coupling of $J=-35$\,K. On top of the CASSCF calculation, containing only static correlations, we add the dynamical correlations via a MRCI calculation. A very efficient truncation of the full MRCI is provided by the DDCI method, which includes only those excitations that actually contribute to energy difference between the spin states, up to second order.\cite{ddci_1} Different types of DDCI methods have been designed and differ by their level of truncation. The most accurate is the DDCI3 method for which even a minimum CAS is sufficient,\cite{cas_ddci_1,*cas_ddci_2} while, e.g., DDCI2 cannot reach that level of accuracy even when applied on an extended CAS.\cite{cas_ddci_2,ddci_3} For the intradimer coupling in callaghanite, DDCI3 yields $J=-66$\,K. The inclusion of Davidson corrections,\footnote{The Davidson term corrects for unlinked quadrupole excitations not considered in MRCI(single,double) type calculations such as DDCI} as suggested by some authors,\cite{dav_corr1,mn2} reduces $J$ to $-45$\,K. We checked the qualitative stability of our results by comparing different truncation levels of DDCI, varying the charges of the embedding as well as the quality of the basis set, and found that the resulting FM character of $J$ is very robust. However, the coupling strength is definitely overestimated compared with our experimental results (see Sec.~\ref{hc}).

\section{Experimental results}
\label{sec:experiment}

\subsection{\label{sample}Sample characterization}
Powder XRD measurements~\cite{supplement} reveal callaghanite with a small admixture (\hbox{$<2$\%}) of hydromagnesite, Mg$_5$(CO$_3$)$_4$(OH)$_2$$\cdot$4H$_2$O, the common accompanying mineral for samples from the Premier Chemical Magnesium Mine. Hydromagnesite is tightly intergrown with callaghanite, hindering a simple mechanical separation of these two phases. However, hydromagnesite is non-magnetic and should thus not affect the magnetic measurements, aside from a slight change of the effective sample mass.

\subsection{\label{magnetization}Magnetization measurements}
\begin{figure}[tb] \includegraphics[width=8.6cm]{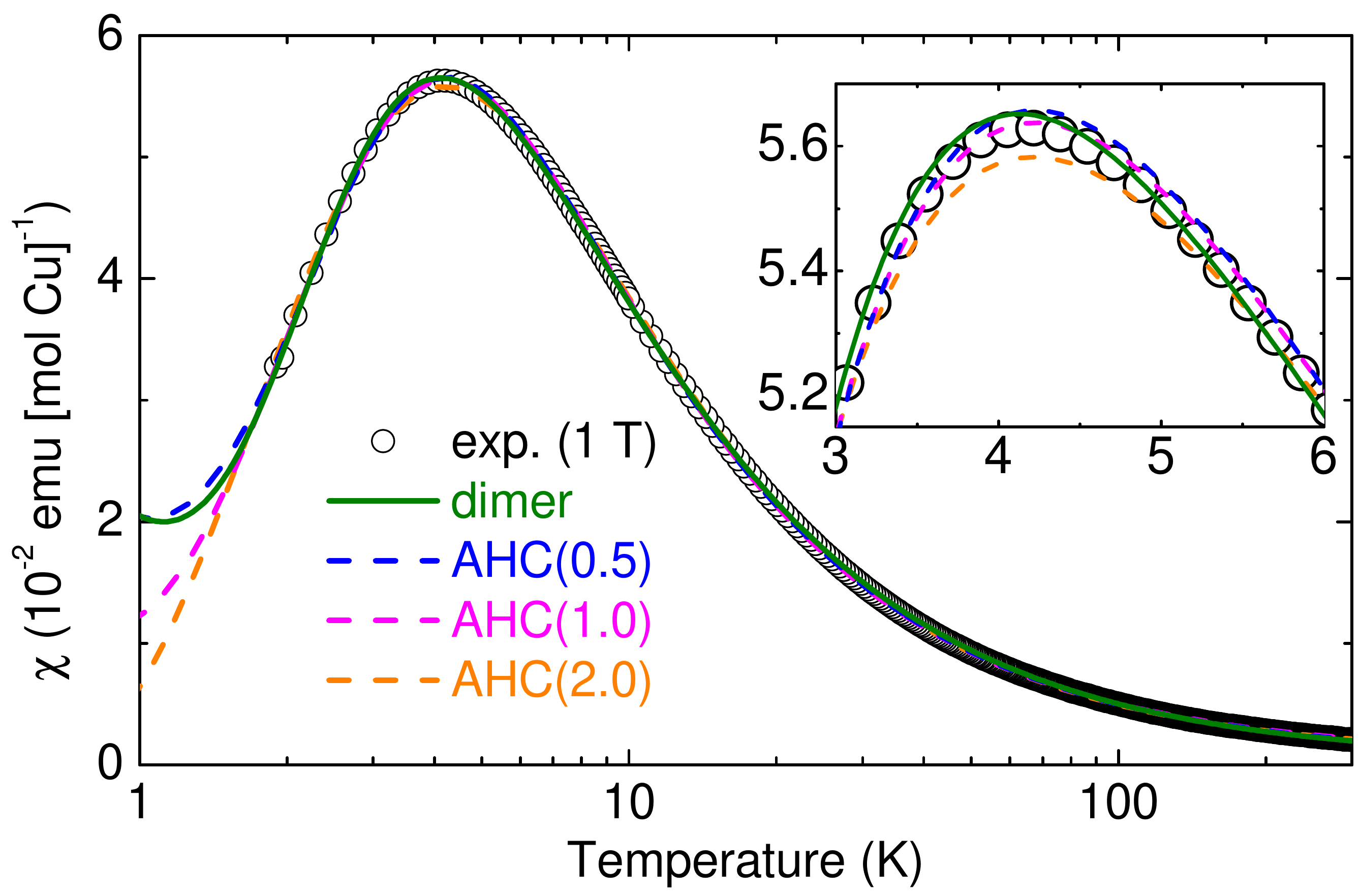}
\caption{\label{sus}(Color online) The experimental susceptibility data collected at a magnetic field of 1\,T and dimer and alternating Heisenberg chain (AHC) fits for different ratios $\alpha=-J_0'/J_0$.}
\end{figure}

The temperature-dependent susceptibility curve $\chi(T)$ is shown in Fig.~\ref{sus}. It features a dome-like peak, which is typical for low-dimensional quantum magnets, with a maximum at about 4\,K and no signatures of magnetic ordering. According to the crystal structure and the results from the TB-analysis (see Sec.~\ref{tb}), a dimer model represents a natural choice for fitting the experimental data. Such a model supplemented by impurity contributions, $\chi=\chi_0+C_{\text{imp}}/T+\chi_{\text{dimer}}$, indeed provides a perfect fit (Fig.~\ref{sus}, Table~\ref{susFit}) with an AFM coupling of $J_0=7$\,K. 

In the following, we use the notation $J_0$ and $J_0'$ for the couplings within and between the AFM spin dimer, respectively. Two options should be considered: i) $J_0=J$ and $J_0'=J'$, i.e., the coupling within the structural dimers is AFM, as evidenced by DFT+$U$ at lower $U_d$; ii) $J_0'=J$ and $J_0=J'$, i.e., the AFM dimers are formed by $J'$ whereas $J$ is FM according to DFT+$U$ at higher $U_d$, PBE0 and WF-based methods. Then the relevant model is the FM-AFM alternating Heisenberg chain (AHC) model.

Indeed, we can reproduce the experimental susceptibility with the ratio of $\alpha=-J_0'/J_0$\,=\,0\,--\,2.0, where $\alpha=0$ corresponds to the isolated dimer model. For about $\alpha>1$, the agreement with the experiment declines, as shown in the inset of Fig.~\ref{sus}. The fitted parameters for selected ratios $\alpha$ are provided in Table~\ref{susFit}. The $g$-values increase with decreasing $\alpha$ and are within the typical range for cuprates.\cite{malachite} The values of $C_{\text{imp}}$ imply less than $5$\% spin-1/2 paramagnetic impurities in our sample. Although both the dimer and AHC models are compatible with the susceptibility data, rather sharp limits on the exchange couplings are obtained. We find that the AFM exchange is about 7\,K, while the FM exchange ranges between 0 and $-8$\,K.

\begin{table}[tbp]
\begin{ruledtabular}
\caption{\label{susFit} 
The parameters obtained by fitting the experimental susceptibility data $\chi(T)$ with a dimer and FM-AFM alternating Heisenberg chain (AHC) models for different ratios $\alpha=-J_0'/J_0$. $\chi_0$ and $C_{\text{imp}}$ are given in (emu/mol) and ($\times 10^{-4}$ emu K/mol), respectively. }
\begin{tabular}{c c c c c c}
model              & $J_0$ (K)   &  $J_0'$ (K) & $g$      & $\chi_0$               & $C_{\text{imp}}$ \\ \hline
dimer              &  7.00       &   --        &  2.19    &  $3.20$   &  0.019    \\
AHC($\alpha=0.2$)  &  7.05       &  $-1.41$    &  2.16    &  $4.83$   &  0.019    \\
AHC($\alpha=0.5$)  &  7.20       &  $-3.60$    &  2.12    &  $6.07$   &  0.017    \\
AHC($\alpha=1.0$)  &  7.35       &  $-7.35$    &  2.11    &  $7.49$   &  0.008    \\
AHC($\alpha=1.25$) &  7.39       &  $-9.24$    &  2.11    &  $7.87$   &  0.002    \\
AHC($\alpha=2.0$)  &  8.03       &  $-16.06$   &  2.10    &  $7.17$   &  0.000    \\
\end{tabular}
\end{ruledtabular}
\end{table}

\begin{figure}[tb] \includegraphics[width=8.6cm]{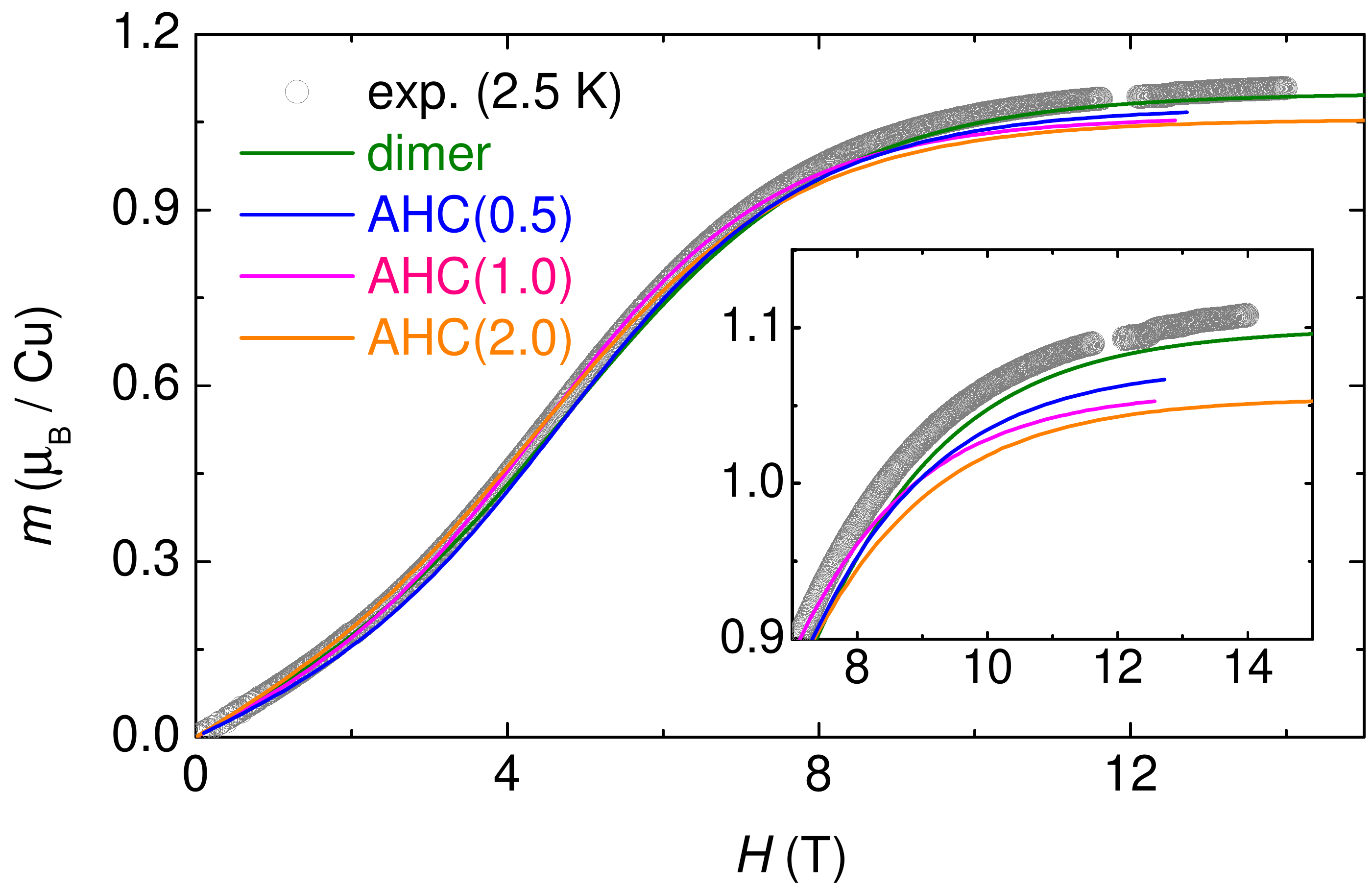}
\caption{\label{mH}(Color online) Field-dependent magnetization data collected at a temperature of 2.5\,K. The labels ``dimer'' and ``AHC'' denote the isolated dimer and alternating Heisenberg chain models, respectively, where the ratio $\alpha$ is given in brackets. The model curves are calculated with the parameters from Table~\ref{susFit}.}
\end{figure}

Field-dependent magnetization $m(H)$ has turned out as a viable tool for distinguishing between different models and might help narrowing the possible range of $\alpha$. Owing to the weak couplings in callaghanite, full saturation of spin-$\frac12$ moments can be reached already at 14\,T, as shown in Fig.~\ref{mH}, where the displayed dimer and AHC curves are calculated with the parameter sets of Table~\ref{susFit}. For fields above 10\,T, differences between the models become most apparent and it is evident that only the dimer model and AHC fits for $\alpha$ considerably smaller than 0.5 can reproduce the experimental data. This also means that the maximum absolute strength of exchange couplings in callaghanite is about 7\,K and, thus, considerably below the computational estimates applying PBE0 and DDCI.

\subsection{\label{hc}Specific heat measurements}
\begin{figure}[tb] \includegraphics[width=8.6cm]{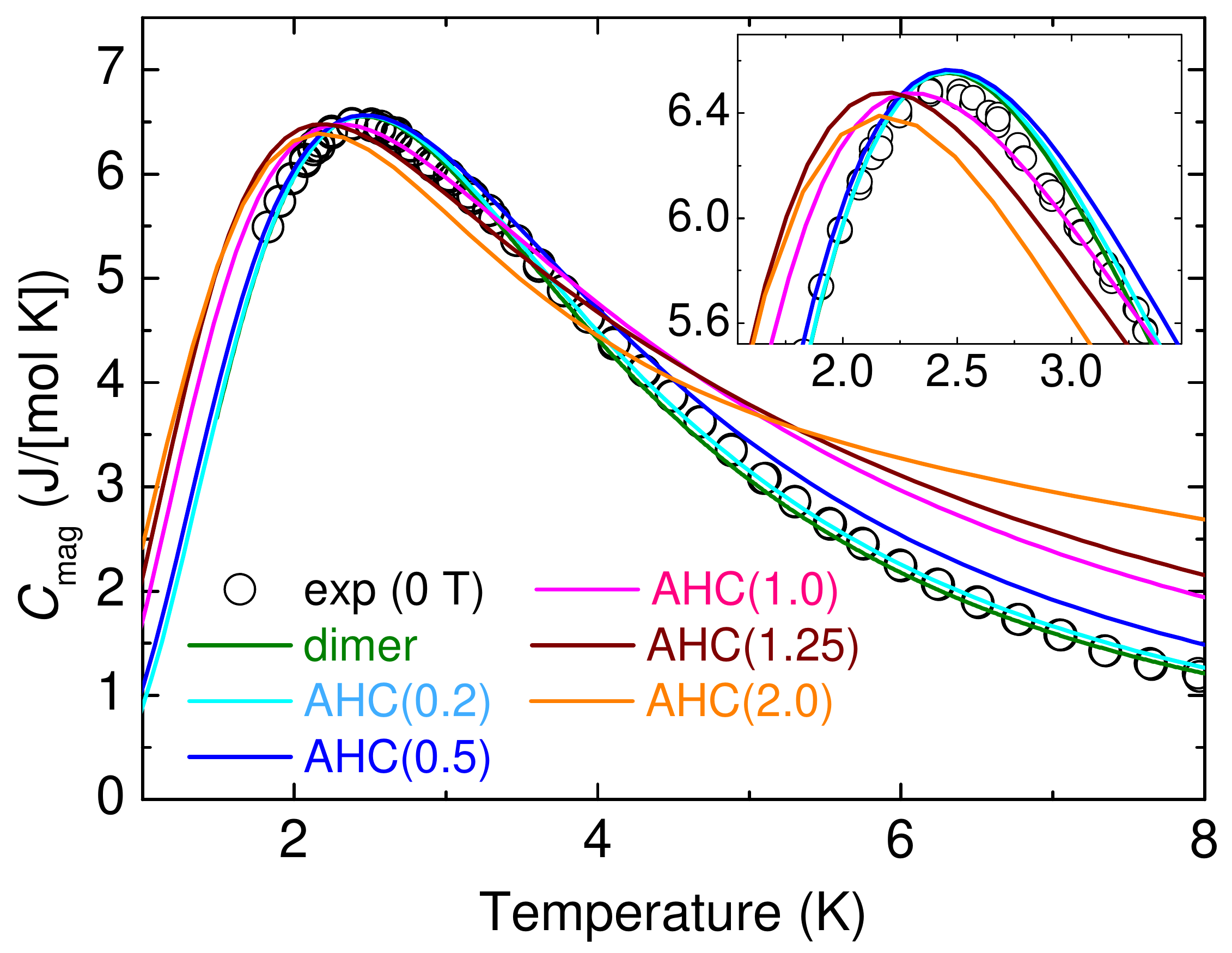}
\caption{\label{cp}(Color online) The magnetic contribution to the specific heat, $C_{\mg}$, at zero magnetic field. The labels ``dimer'' and ``AHC'' denote the isolated dimer and alternating Heisenberg chain models, respectively, where the ratios $\alpha$ are given in brackets. The model curves are calculated with the parameters from Table~\ref{susFit}.}
\end{figure}

\begin{figure}[tb] \includegraphics[width=8.6cm]{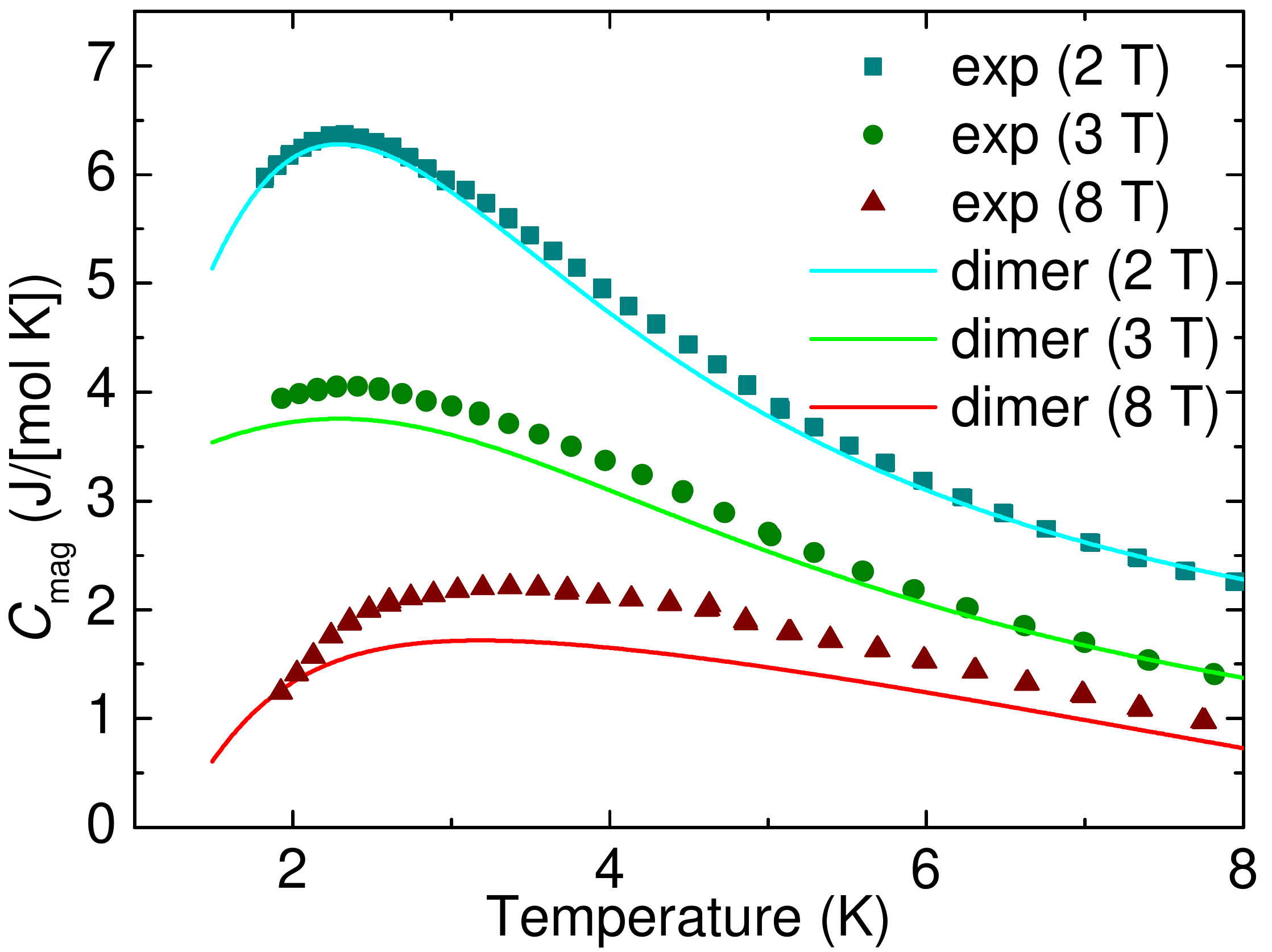}
\caption{\label{cpH}(Color online) The magnetic contribution to the specific heat, $C_{\mg}$, for magnetic fields of 2, 3 and 8\,T. Solid lines represent the results of the dimer model. For reasons of clarity, the 2\,T and 8\,T curves are shifted by $+1$ and $-2$\,J/(mol K), respectively. The model curves are calculated with the parameters from Table~\ref{susFit}. Results for the AHC model with $\alpha\leq0.2$ are almost indistinguishable from the dimer curves and, thus, not displayed.}
\end{figure}

As a next step, specific heat data are collected in a temperature range from 1.8--40\,K for magnetic fields 0--8\,T. At 0\,T, the curve features a peak at 2.38\,K (Fig.~\ref{cp}). The field-dependence of this peak pinpoints its magnetic origin. The application of a magnetic field suppresses the maximum and increases the peak width (see Fig.~\ref{cpH} and supplemental material~\cite{supplement}). For $H<2$\,T, the peak shifts to lower temperatures while above 2\,T a shift towards higher temperatures is observed. Above 4\,T, additionally, the peak value starts to increase. 

In order to analyze the magnetic contribution to the specific heat $C_{\mg}$, the lattice part should be subtracted first. However, we were unable to fit the lattice part with a single Debye function or even their linear combination. This problem is probably caused by a very complex crystal structure of callaghanite, where localized vibration modes not described by the Debye model are expected. Therefore, we used a simplified approach and performed an empirical fitting of the experimental heat-capacity data above 11\,K with a third-order polynomial augmented by the $A/T^2$ term that accounts for the high-temperature limit of $C_{\mg}$ (Ref.~\onlinecite{johnston2000}, see also the supplemental material~\cite{supplement}). The reliability of this procedure was checked by integrating $C_{\mg}/T$ and evaluating the magnetic entropy that amounts to about 80\% of the theoretical value of $R\ln 2$ for spin-1/2. The remaining discrepancy can be attributed to impurity contributions and to a systematic experimental error caused by the very small available sample size.

Owing to the narrow peak width, the zero-field data are best suited for a comparison between the dimer and AHC models with the parameters fixed to those of Table~\ref{susFit}. The dimer model allows for an accurate description of $C_{\mg}$ (Fig.~\ref{cp}). A similar good agreement can only be obtained for AHC models with $\alpha\leq0.2$, i.e. with extremely weak interdimer coupling of $|J'|<1.5$\,K. The evolution of $C_{\mg}$ in a magnetic field is also nicely reproduced with these models (Fig.~\ref{cpH}). For fitting the $C_{\mg}$ data, the AHC and dimer functions had to be scaled down by about 20\% in order to account for the too low height of the experimental magnetic peak. This downscaling compensates for the missing magnetic entropy. Despite these technical difficulties, the specific-heat data clearly evidence that callaghanite features magnetic dimers with an AFM coupling of $J_0\approx7$\,K. FM interdimer couplings, if present at all, are very weak. Their absolute values are below 1.5\,K.

\section{Discussion}
\label{sec:discussion}
In the present study, we discuss the magnetic properties and the microscopic magnetic model for the rare Cu$^{2+}$-mineral callaghanite. Its crystal structure consists of isolated Cu$_2$(OH)$_6$ dimers with Mg$^{2+}$-ions as well as loosely bonded carbonate groups and water molecules in between. Susceptibility, field-dependent magnetization and specific heat data can be described by a model of AFM dimers featuring a weak exchange coupling $J_0$ of about 7\,K. Very weak FM couplings $J_0'<|-1.5|$\,K between the dimers might be present as well. Intuitively, the structural and magnetic dimers might be equated, but such guessing of magnetic models is, in general, misleading as shown for many other dimer compounds.\cite{clinoclase,malachite,vodpoh} 

Experimentally, thermodynamic measurements are usually unable to determine the position of magnetic dimers in the crystal structure, because these measurements are sensitive only to the topology of the spin lattice. More elaborate experiments, such as inelastic neutron scattering providing a $q$-dependent probe of the magnetic system, would be required to determine the position of magnetic dimers experimentally. Alternatively, computational techniques can be used to determine the strongest AFM coupling in the system and, thus, ascribe the spin dimer to a certain exchange pathway in the crystal structure. Unfortunately, magnetic couplings in callaghanite are so weak that they are difficult to evaluate from first principles. We were unable to determine whether $J$ is ferromagnetic or antiferromagnetic, but, fortunately, the computational results for $J'$ are unambiguous and yield the upper estimate of $J'=2.2$\,K, which is more than three times smaller than $J_0\simeq 7$\,K. Therefore, $J'$ can not be responsible for the formation of AFM spin dimers. Then $J=J_0$, hence the magnetic and structural dimers coincide.

The formation of weak AFM spin dimers on the structural Cu$_2$(OH)$_6$ dimers may have interesting implications for further experiments, because even a weak external pressure will change the Cu--O--Cu bridging angle and, therefore, will have significant impact on the intradimer coupling. A particularly interesting situation may arise if the AFM coupling is reduced under pressure and eventually becomes ferromagnetic, thus leading to a peculiar pressure-induced phase transition. Moreover, it will render callaghanite an excellent model system for studying the subtle balance between FM and AFM couplings close to compensation. In fact, even the ambient-pressure behavior of callaghanite is a challenge for computational methods that are unable to determine the sign of $J$ unambiguously. 

The conventional DFT+$U$ approach yields both FM and AFM $J$ depending on the Coulomb repulsion $U_d$. Although experimental data can be used to fine-tune the value of $U_d$ and obtain the correct AFM $J$ of about 7\,K, this strategy is hardly acceptable, because it renders the computational approach essentially empirical. Therefore, we tried to use alternative computational techniques that, albeit more demanding, should be free from adjustable parameters. With PBE0, we got a FM $J$ of about $-25$\,K, where the sign of the coupling is wrong, and the strength of the coupling is far too large compared to the experimental energy scale. As demonstrated in Sec.~\ref{ssec:hybrids}, exchange constants calculated within PBE0 may largely deviate from the experiment. Therefore, the large error for the very weak couplings in callaghanite is not unexpected.

The DDCI3 method, designed for calculating energy differences and considered as one of the most accurate methods available for such purposes,\cite{hybrid2,ddci_acc2,chem_rev} was employed together with a size-converged basis set of triple-zeta quality. For the intradimer coupling, we got $J=-66$\,K that reduced to $-45$\,K by including Davidson-corrections, where in both cases the deviations from experimental data are even larger than for PBE0 (Table~\ref{J_cond}). Regarding the proven track record of DDCI3,\cite{hybrid2,ddci_acc2,chem_rev} this appears surprising and may be related to the low (monoclinic) symmetry of callaghanite. The low symmetry dramatically increases the computational effort and restricts the cluster size. Additionally, convergence problems occur way more easily. In fact, almost all periodic compounds studied so far with this method have at least orthorhombic and, typically, even higher symmetry. A further source for inaccuracies and ambiguities arises from certain corrections (e.g. Davidson corrections), which we have shown to change the results considerably. Eventually, details in the embedding, which are scarcely discussed in the literature, play an important role for reaching good agreement between computational results and experiment. 

\begin{table}[tbp]
\begin{ruledtabular}
\caption{\label{J_cond} 
The intra- and interdimer exchanges, $J$ and $J'$, respectively, from experiments and calculated with the different theoretical methods. For LSDA*$U$, $U_d=6.5\pm0.5$ and $U_d=8.5\pm1.0$ are used for AMF and FLL DCCs, respectively. DDCI3+Q denotes DDCI3 results including Davidson corrections. While the position of $J$ and $J'$ in the crystal structure is defined for the theoretical results it is unknown for the experiment, thus, we distinguish between $J$ and $J_0$ as well as $J'$ and $J_0'$. $J_0'$ is FM, i.e. $<0$. }
\begin{tabular}{c c c}
method           &  $J$         &  $J'$          \\ \hline
experiment       &  $J_0=7$     &  $|J_0'|<1.5$ \\
LSDA+$U$ (AMF)   &  $-12\pm20$  &  $1.6\pm 0.1$  \\ 
LSDA+$U$ (FLL)   &  $35\pm35$   &  $2.0\pm 0.2$  \\
PBE0             &  $-25$       &    --          \\
UHF              &  $-48$       &    --          \\
CASSCF           &  $-35$       &    --          \\
DDCI3            &  $-66$       &    --          \\
DDCI3+Q          &  $-45$       &    --          \\
\end{tabular}
\end{ruledtabular}
\end{table}

Given the low accuracy of hybrid functionals and WF-based methods for the weak couplings in callaghanite, the results of the DFT+$U$ methods deserve a closer examination. Remarkably, they show a sizable ambiguity for $J$ and nearly no ambiguity for $J'$, even though $J'$ is weaker than $J$. We argue that this effect is related to the different mechanisms of these couplings. The intradimer coupling $J$ runs between the two edge-sharing CuO$_4$ plaquettes. It includes sizable FM and AFM contributions, as typical for Cu--O--Cu bridging angles close to $90^{\circ}$. The interdimer coupling $J'$ connects two CuO$_4$ plaquettes having no common oxygen atoms. This coupling is of super-superexchange type (Cu--O$\ldots$O--Cu or even more extended pathways) and features a predominant AFM contribution. The DFT+$U$ methods are quite efficient in reproducing even very weak couplings of the latter type, where only the AFM term is relevant. In contrast, the compensation of large FM and AFM terms in the couplings of the former type remains challenging for computational methods (see also Ref.~\onlinecite{cdvo3}, where a similar analysis for V$^{4+}$ oxides has been performed). We have shown that the DFT+$U$ results on the sign of $J$ are inconclusive, whereas hybrid functionals and WF-based methods yield FM $J$ contradicting the experiment.

The application of DFT+$U$ to the evaluation of magnetic couplings requires a careful choice of $U_d$ and other computational parameters to avoid ambiguity and obtain correct estimates of $J$'s. This strategy has proved to be very efficient, yet it has its limitations for weak couplings, where even minor changes in $U_d$ lead to large ambiguities in the resulting exchange couplings. Considering our results for callaghanite, we conclude that even the couplings of $2-3$\,K can be calculated unambiguously as long as these couplings are of super-superexchange type. In contrast, the evaluation of direct-exchange and superexchange involves larger ambiguities. Here, only the couplings of $20-30$\,K are obtained unambiguously in the sense that their sign is safely established by DFT+$U$. We hope that this analysis will be a useful guidance for future computational work on magnetic couplings in insulators.

\section{Summary}
A combined theoretical and experimental study of the Cu$^{2+}$-compound callaghanite is presented. The crystal structure of this mineral features well isolated Cu$_2$(OH)$_6$ dimers exhibiting a Cu--O--Cu bridging angle of about $96^{\circ}$. Therefore, according to common empirical rules, an intradimer exchange coupling close to compensation, i.e. the transition from the ferromagnetic to the antiferromagnetic regime, can be expected. Indeed, susceptibility, field-dependent magnetization and specific heat measurements reveal a very small spin gap of about 7\,K where all the experimental data can be interpreted within an isolated-dimer model. FM interactions between the magnetic dimers cannot be excluded but are below $|-1.5|$\,K. 

Since the experimental results do not provide the actual position of the magnetic dimer in the crystal structure, DFT+$U$ calculations were employed yielding a coupling strength close to zero for the long-range interdimer coupling. This provides clear evidence that magnetic and structural dimers are the same. By contrast, estimates for the weak short-range intradimer coupling left ambiguities which could not be resolved even by employing PBE0 and highly elaborating wave-function based DDCI3 methods. 

Reasons for the different performance with respect to characteristics of the exchange pathways were analyzed and minimum coupling strengths required for qualitatively reliable results were discussed. With respect to the small energy scales in callaghanite, which are fully accessible with experimental techniques, we emphasized the possibility of interesting high-pressure physics in callaghanite which will be addressed in a future study.
\label{sec:summary}

\acknowledgments
We acknowledge the experimental support by Yurii Prots and Horst Borrmann (laboratory XRD), Carolina Curfs (ID31) and the provision of the ID31 beamtime by the ESRF. We are grateful to Yurii Skourski for high-field magnetization measurements. We would also like to thank the Department of Materials Research and Physics of the Salzburg University for providing the natural sample of callaghanite from their mineralogical collection (inventory number 9107) and G. J. Redhammer for fruitful discussions. We further thank B. Paulus for helpful suggestions concerning WF-based calculations. AT and OJ were supported by the Mobilitas program of the ESF (grant numbers MTT77 and MJD447). SL acknowledges the funding from the Austrian Fonds zur F\"orderung der wissenschaftlichen Forschung (FWF) via a Schr\"odinger fellowship (J3247-N16).


%

\clearpage

\begin{table*}[H]
\begin{tabular}{c}
\huge{\texttt{Supporting Material}} \\
\end{tabular}
\end{table*}

\begin{widetext}

\begin{table}[H]
\begin{ruledtabular}
\caption{\label{H_opt} 
The fractional coordinates of hydrogen as obtained from a GGA optimization of the atomic H-positions. The lattice parameters and all other atomic positions were fixed to those of the room temperature single crystal XRD structure. The convergency criterion for remaining forces was set to 1\,meV/\r{A}.}
\begin{tabular}{c c c c}
atom & $x/a$ & $y/b$ & $z/c$ \\ \hline
H1 & -0.2615   &  0.0932   &  0.13923   \\
H2 & -0.1206   &  0.2470   & -0.26084   \\ 
H3 & -0.0288   & -0.2982   &  0.00571   \\ 
H4 &  0.1484   &  0.1215   & -0.15828   \\ 
H5 & -0.2088   & -0.0497   & -0.08354   \\ 
\end{tabular}
\end{ruledtabular}
\end{table}

\begin{figure}[H]
\includegraphics[width=10cm]{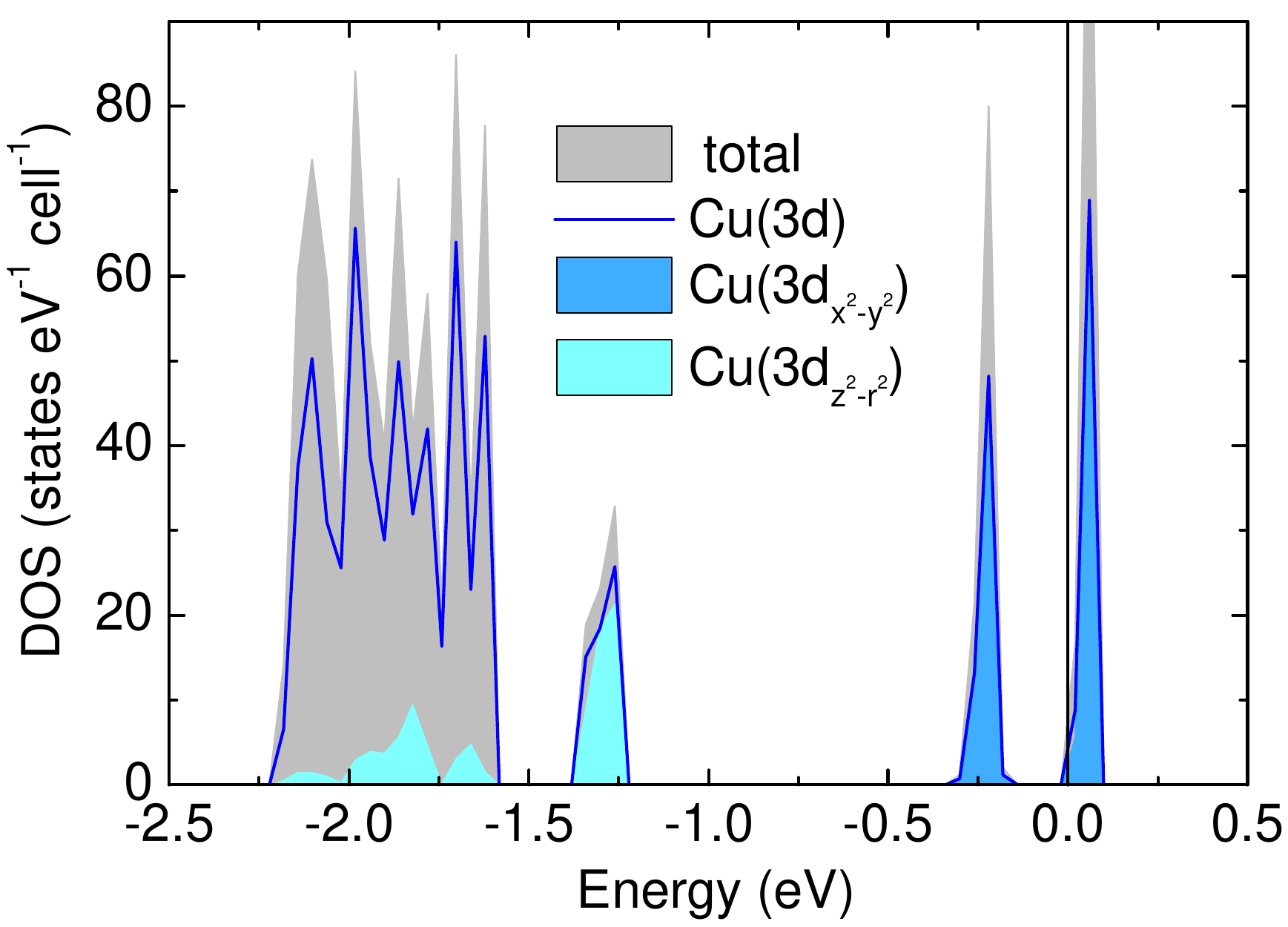}
\caption{\label{dos_orb}
(Color online) The total and orbital resolved density of states (DOS) near the Fermi level. Cu(3$d_{x^2-y^2}$) and Cu(3$d_{z^2-r^2}$) denote the partial DOSs of the respective Cu(3$d$) orbitals. Cu(3$d$) gives the total Cu(3$d$) DOS.}
\end{figure}

\begin{table} [H]
\caption{
Refined atomic positions (in fractions of lattice parameters) and isotropic atomic displacement parameters $U_{\text{iso}}$ (in $10^{-2}$~\r A$^2$) for the callaghanite structure at 10\,K (first row) and at room temperature (second row). C and O2 are in the $4e$ position, and all other atoms are in the general position $8f$ of the space group $C2/c$. Lattice parameters are as follows: $a=9.98324(2)$~\r A, $b=11.75057(2)$~\r A, $c=8.16740(2)$~\r A, $\beta=107.3731(2)^{\circ}$ at 10~K ($R_I=0.034$) and $a=10.01079(3)$~\r A, $b=11.75583(3)$~\r A, $c=8.21646(1)$~\r A, $\beta=107.3968(2)^{\circ}$ at room temperature ($R_I=0.044$). Hydrogen positions were not refined. All standard deviations refer to the Rietveld refinement, only.
}

\begin{minipage}{15cm}
\begin{ruledtabular}
\begin{tabular}{ccccc}
  Atom & $x/a$      & $y/b$      & $z/c$     & $U_{\text{iso}}$ \\
  Cu  & 0.04870(8)  & 0.10832(7) & 0.45651(9)  & 0.27(2)          \\
      & 0.04890(10) & 0.10788(9) & 0.45648(12) & 0.97(2)          \\
  Mg  & 0.1568(2)   & 0.3152(2)  & 0.3277(2)   & 0.18(5)          \\
      & 0.1574(3)   & 0.3151(2)  & 0.3287(3)   & 1.1(1)          \\
  C   & 0.0         & 0.5427(7)  & 0.25        & 0.9(2)          \\
      & 0.0         & 0.5408(10) & 0.25        & 1.6(3)          \\
  O1  & 0.1183(3)   & 0.4861(3)  & 0.2773(4)   & 0.02(4)          \\
      & 0.1175(5)   & 0.4860(4)  & 0.2787(6)   & 1.5(1)          \\
  O2  & 0.0         & 0.6522(4)  & 0.25        & 0.02(4)          \\
      & 0.0         & 0.6511(6)  & 0.25        & 1.9(2)          \\
  O3  & 0.0183(4)   & 0.2651(3)  & 0.0954(4)   & 0.02(4)          \\
      & 0.0180(4)   & 0.2647(3)  & 0.0948(5)   & 0.37(12)          \\
  O4  & 0.1175(3)   & 0.9512(3)  & 0.4954(4)   & 0.02(4)          \\
      & 0.1177(4)   & 0.9503(4)  & 0.4982(6)   & 0.74(13)          \\
  O5  & 0.2249(3)   & 0.1519(3)  & 0.4125(4)   & 0.02(4)          \\
      & 0.2251(5)   & 0.1512(3)  & 0.4150(6)   & 0.45(12)          \\
  O6  & 0.3367(3)   & 0.3277(3)  & 0.2475(4)   & 0.02(4)           \\
      & 0.3351(4)   & 0.3269(3)  & 0.2442(5)   & 0.79(13)           \\

\end{tabular}
\end{ruledtabular}
	  
\end{minipage}
\end{table}

\clearpage

\begin{figure}
\includegraphics[width=10cm]{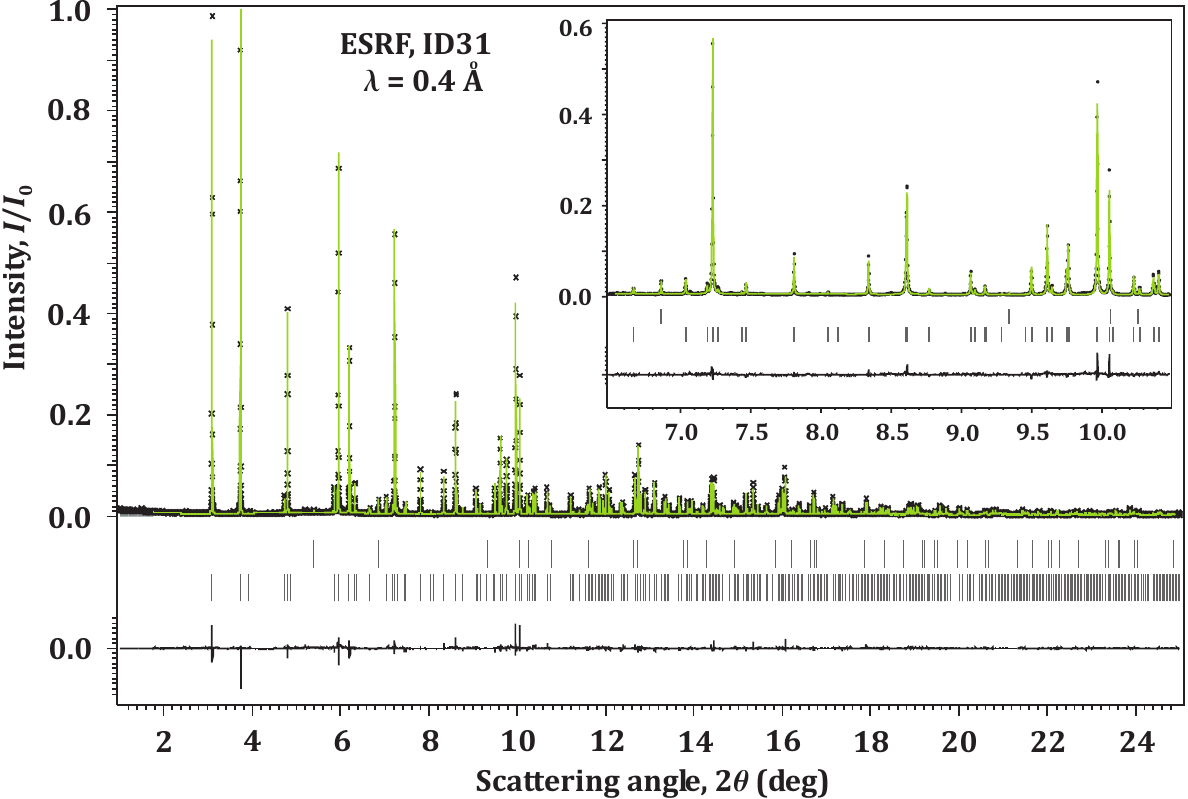}
\caption{
Rietveld structure refinement for callaghanite at room temperature. Upper and lower ticks show reflections positions for the quartz (0.05\,wt.\%) and callaghanite (99.95\,wt.\%) phases. Few reflections of the hydromagnesite impurity were relatively broad and nearly invisible and, therefore, not included in the refinement.}
\end{figure}

\begin{figure}[h]
\includegraphics[width=15cm]{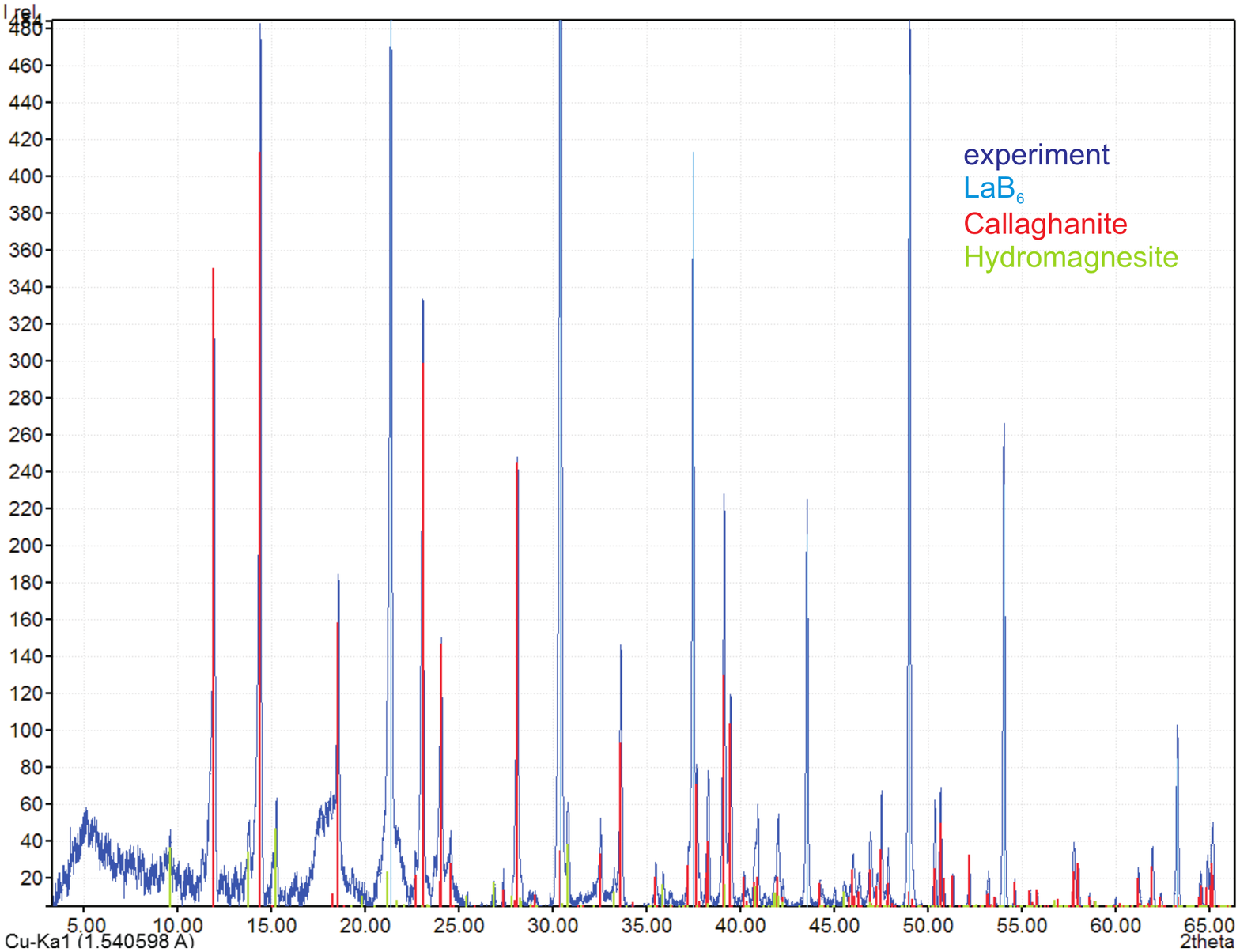}
\caption{\label{xrd}
(Color online) Room temperature powder x-ray diffraction pattern (Huber G670 Guinier camera, CuK$_{\alpha\,1}$ radiation, ImagePlate detector, $2\theta\,=\,3-100^{\circ}$ angle range) of the callaghanite sample bearing some impurities from hydromagnesite.}
\end{figure}

\begin{figure}[h]
\includegraphics[width=10cm]{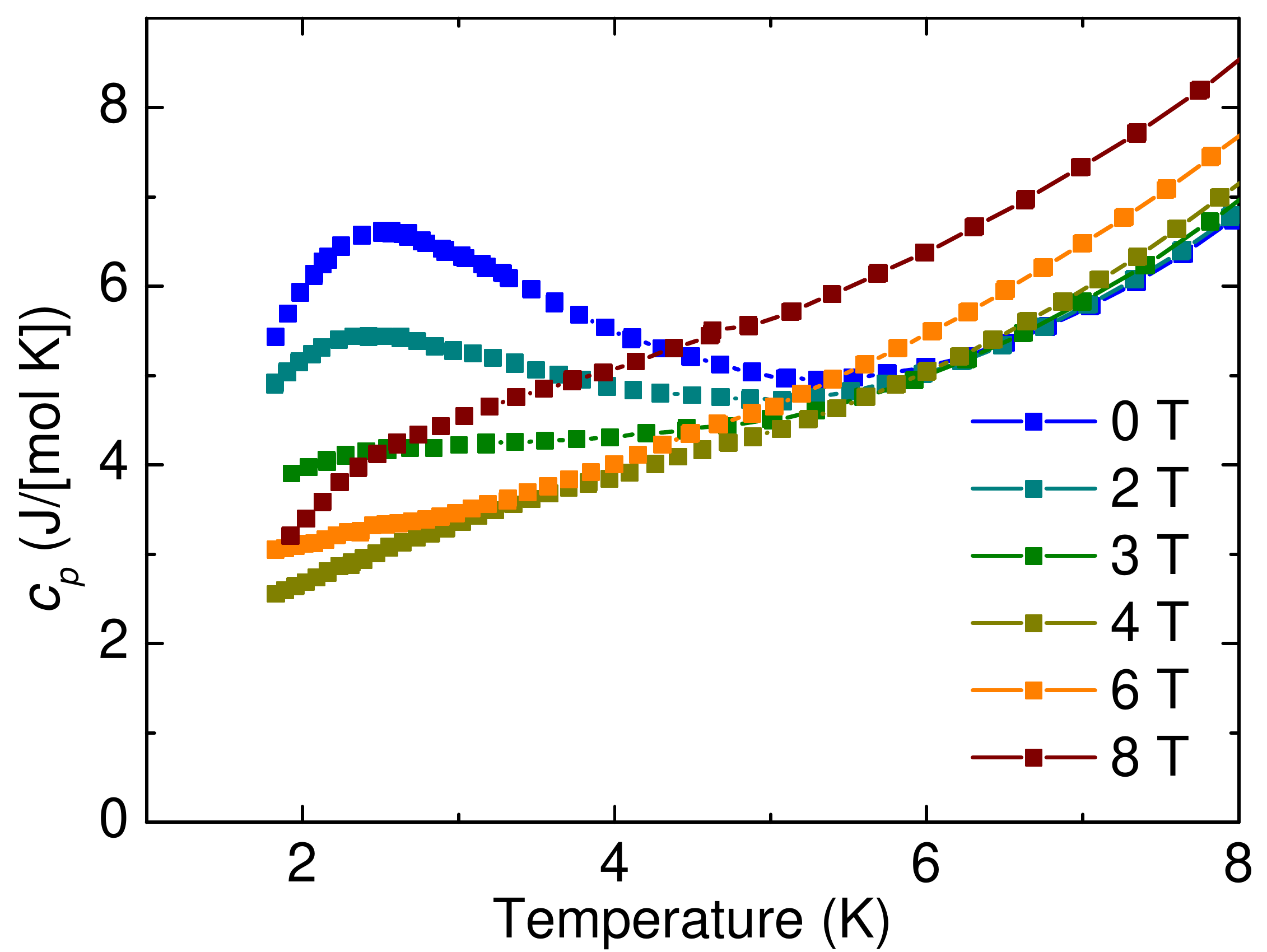}
\caption{\label{cp_bg}
(Color online) Specific heat data of callaghanite collected in magnetic fields up to 8\,T.}
\end{figure}

\begin{figure}[h]
\includegraphics[width=10cm]{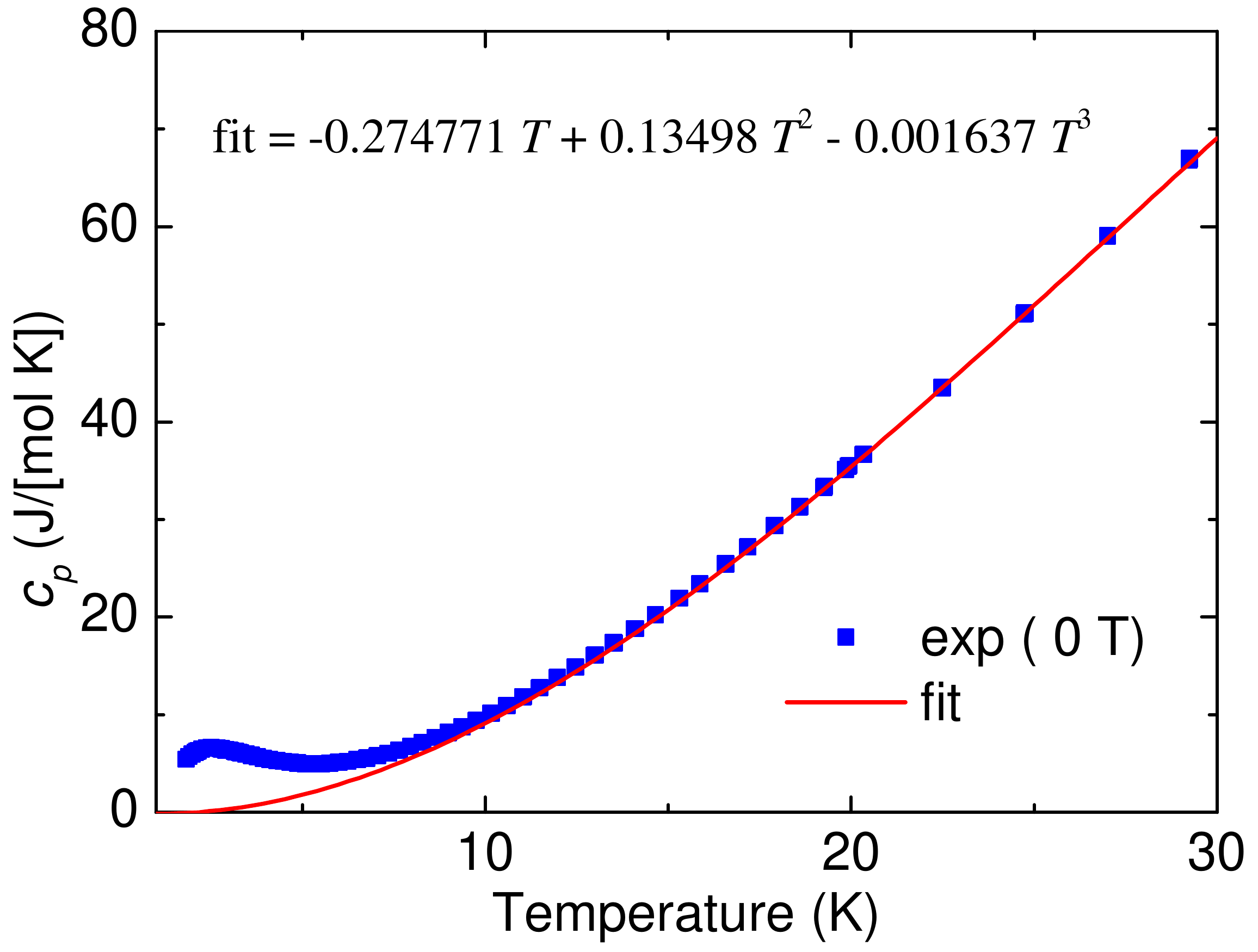}
\caption{\label{cp_bg}
(Color online) Specific heat data of callaghanite collected in zero magnetic field. The red line shows the fit with an arbitrary background polynomial of the form $\alpha \cdot T + \beta \cdot T^2 + \gamma \cdot T^3$, which we subtracted to get the magnetic contribution to the specific heat $C_{\text{mag}}$.}
\end{figure}

\bigskip

\end{widetext}

\end{document}